\documentclass[11pt]{article}
\usepackage{latexsym}
\usepackage{amsmath}
\usepackage{amssymb}
\usepackage[dvips]{graphicx}
\usepackage{latexsym}

\newcommand{\C}{\mathbb{C}}

\newcommand{\spS}{\mathbb{S}}
\newcommand{\Z}{\mathbb{Z}}
\newcommand{\T}{\mathbb{T}}
\newcommand{\F}{\mathbb{F}}
\newcommand{\M}{\mathbb{M}}

\newcommand{\hook}{{\setlength{\unitlength}{11pt}   
                   \begin{picture}(.833,.8)
                   \put(.15,.08){\line(1,0){.35}}
                   \put(.5,.08){\line(0,1){.5}}
                   \end{picture}}}

\def\p{\partial}
\def\Kww{K_{w \bar{w} }}
\def\Kwz{K_{w \bar{z} }}
\def\Kzw{K_{z \bar{w} }}
\def\Kzz{K_{z \bar{z} }}
\def\Kw{K_w}
\def\Kwbar{K_{\wbar}}
\def\wbar{\bar{w}}
\def\zbar{\bar{z}}

\begin{document}

\title{\vskip -70pt
\begin{flushright}
{\normalsize DAMTP-2012-39} \\
\end{flushright}
\vskip 10pt
{\bf Quaternion-K\"ahler four-manifolds and Przanowski's function 
\vskip 15pt}}
\author{Moritz H\"ogner\thanks{Email: M.Hoegner@damtp.cam.ac.uk}
\\
Department of Applied Mathematics and Theoretical Physics,\\
University of Cambridge,\\
Wilberforce Road, Cambridge CB3 0WA, UK. }
\date{}
\maketitle

\begin{abstract}
Quaternion-K\"ahler four-manifolds, or equivalently anti-self-dual Einstein manifolds, are locally determined by one scalar function subject to Przanowski's equation. Using twistorial methods we construct a Lax Pair for Przanowski's equation, confirming its integrability. The Lee form of a compatible local complex structure, which one can always find, gives rise to a conformally invariant differential operator acting on sections of a line bundle. Special cases of the associated generalised Laplace operator are the conformal Laplacian and the linearised Przanowski operator. We provide recursion relations that allow us to construct cohomology classes on twistor space from solutions of the generalised Laplace equation. Conversely, we can extract such solutions from twistor cohomology, leading to a contour integral formula for perturbations of Przanowski's function. Finally, we illuminate the relationship between Przanowski's function and the twistor description, in particular we construct an algorithm to retrieve Przanowski's function from twistor data in the double-fibration picture. Using a number of examples, we demonstrate this procedure explicitly.
\end{abstract}

\section{Introduction}
\subsection{Motivation}
Quaternion-K\"ahler four-manifolds can be characterised in three different ways. Firstly, with motivation coming from higher-dimensional Quaternion-K\"ahler manifolds which are Riemannian manifolds with holonomy $Sp(n) Sp(1)$, one can define a four-dimensional Quaternion-K\"ahler manifold to be anti-self-dual Einstein with non-vanishing cosmological constant \cite{SalamonQKMfld}. Secondly, as is well-known, they can be described by means of their twistor space, a three-dimensional complex manifold with a four-parameter family of holomorphic curves and a contact structure \cite{WardSDC, WardWells, Mason, Penrose}. Finally, these manifolds are locally determined by one scalar function, known in the literature as Przanowski's function, which is subject to a second-order partial differential equation \cite{Przanowski}.\\
While this last description is only local in nature, it appears to be very useful in applications, as explicit expressions for the metric in local coordinates are easily obtained. In particular the hypermultiplet moduli space in string theory is an example of a Quaternion-K\"ahler four-manifold, and Przanowski's function has been used in that context \cite{Alexandrov:2009vj, Alexandrov:2006hx, Looyestijn:2008pg}.\\
The purpose of this paper is to introduce Przanowski's function and the associated partial differential equation as well as its linearisation and clarify their geometric origin in the twistor construction. In particular, we construct a Lax Pair for Przanowski's Equation and exhibit its linearisation as the generalised Laplacian associated to a natural, conformally invariant differential operator. We relate solutions of the generalised Laplace equation to twistor cohomology using recursion relations, leading to a contour integral formula for perturbations of Przanowski's function. Eventually, we provide an algorithm that extracts Przanowski's function from twistor data in the double-fibration picture, extending work of \cite{Alexandrov:2009vj}.

\subsection{Summary of main results}
The starting point of this paper is Przanowski's observation \cite{Przanowski} that locally every anti-self-dual Einstein four-manifold $(M,g)$ admits a compatible complex structure and the metric is of the form

\begin{align}
 \label{metric}
g = 2 \left( \Kww dw d\wbar + \Kwz dw d\zbar + \Kzw dz d\wbar + \left( \Kzz + \frac{2}{\Lambda} e^{\Lambda K} \right) dz d\zbar \right),
\end{align}
where $(w,z)$ are holomorphic coordinates, $(\wbar, \zbar)$ their complex conjugates, $\Lambda \neq 0$ is the cosmological constant and $K(w,\wbar, z, \zbar)$ is a real function on $M$ with $K_{w} = \frac{\p K}{\p w}$ and so forth. \eqref{metric} is an ASD Einstein metric if and only if $K$ satisfies Przanowski's equation,

\begin{align}
 \label{Prz}
 \Kzw \Kwz - \Kww \left( \Kzz + \frac{2}{\Lambda} e^{\Lambda K} \right) + K_w K_{\wbar} e^{\Lambda K} =0.
\end{align}
Our three main results are as follows:\\ \vskip-1em
First, we construct a Lax Pair for Przanowski's equation. As we shall see, the vector fields\footnote{The function $\tilde{K}$ is defined in \eqref{tildeK}.}

\begin{align}
\label{LaxPair}
 l_0 &=\p_w - \xi \frac{ \Kwz }{\tilde{K}}  \p_{\wbar} + \xi \frac{ \Kww}{\tilde{K}}  \p_{\zbar}  + \left( \frac{  \p_w \tilde{K} + e^{\Lambda K} \Kw \Kww }{\tilde{K}} - \frac{\Kww}{ \Kwbar } \right) \xi \p_{\xi},\\
\nonumber
 l_1 &=  \p_z - \frac{\xi}{\tilde{K}} \left( \Kzz +\frac{2}{\Lambda} e^{\Lambda K} \right) {\p}_{\wbar} + \xi \frac{ \Kzw}{\tilde{K}} \p_{\zbar}  + \left( \frac{ \p_z \tilde{K} + e^{\Lambda K} \Kw \Kzw - e^{\Lambda K} \Kwbar \xi }{\tilde{K}}  - \frac{\Kzw}{\Kwbar} +  \frac{\xi}{ \Kw}  \right) \xi \p_{\xi}
\end{align}
commute if and only if \eqref{Prz} is satisfied. Here $\xi$ is an auxiliary parameter whose geometric origin lies in the twistor construction where $\xi$ is the coordinate along a twistor line. Thus Przanowski's equation provides another example of an integrable equation coming from anti-self-duality equations in four dimensions. While the twistorial background of this Lax Pair will be explained in detail later on, the advantage of having a Lax Pair for \eqref{Prz} is that it makes many of the usual properties of integrable systems manifest without resorting to the twistor construction explicitly.\\ \vskip-1em

Secondly, for $m \in \Z$ we construct a conformally invariant differential operator $D = d - \frac{m}{2} B$, where $B$ is the Lee one-form of the chosen complex structure on $M$. $D$ acts on sections of the line bundle $L^{m} = \left( \Lambda^{(2,2)}M \right)^{\frac{m}{4}}$. The generalised Laplacian $\ast D \ast D$ reduces to the conformal Laplacian for sections of $L^{-1}$, while for sections of $L^1$ we reproduce a multiple of the linearised Przanowski operator,

\begin{align}
\label{linPrz}
dPrz &= \Kzw \p_{w \zbar} + \Kwz \p_{z \wbar} - \Kww \p_{z \zbar} - \left( \Kzz + \frac{2}{\Lambda} e^{\Lambda K} \right) \p_{w \wbar}\\
\nonumber &+ e^{\Lambda K} \left( \Kw \p_{\wbar} + \Kwbar \p_w + \Lambda \Kw \Kwbar - 2 \Kww \right).
\end{align}
An associated recursion relation allows us to construct cohomology classes in $H^1 \left( T, \mathcal{O}(2m) \right)$ on twistor space $T$ from solutions of the generalised Laplace equation. Conversely, cohomology classes on $T$ provide solutions to this equation, in particular we obtain a contour integral formula for perturbations $\delta K$ of Przanowski's function from $\varPsi \in H^1 \left( T, \mathcal{O}(2) \right)$,

\begin{align}
\label{contint}
 \delta K &=  \frac{1}{2 \pi i} \oint\limits_{\Gamma} \frac{ \varPsi \xi_{A'} d \xi^{A'} }{\left( \xi^{0'} \right)^2 \left( \xi^{1'} \right)^2}
\end{align}
We integrate along a contour $\Gamma$ around the equator of every twistor line with homogeneous coordinates $(\xi^{0'}, \xi^{1'})$.\\ \vskip-1em

Finally, we provide an algorithm to extract Przanowski's function $K$ along with a complex structure on $M$ from twistor data. As a by-product, we deduce the second-order partial differential equation \eqref{Prz}. A similar method has been established in \cite{Alexandrov:2009vj}, however their procedure requires knowledge of the K\"ahler potential and metric on twistor space, while ours doesn't. The steps of our procedure are:\\
\begin{itemize}
 \item Find canonical coordinates $(x,y,t)$ for the contact form $\tau = dx - y dt$ on twistor space $T$ and pull them back to correspondence space $F$.
 \item $M$ inherits a complex structure from the holomorphic surface $S = \{ p \in F \; | \; t = 0 \}$ with holomorphic coordinates $w = y\Big|_S$, $z = x\Big|_S$ and complex conjugates $\wbar = \iota (y) \Big|_{\iota(S)}$, $\zbar = \iota (x) \Big|_{\iota(S)}$, where $\iota$ is the anti-holomorphic involution on twistor space preserving the real lines. The asymmetry of $w$ and $z$ in \eqref{Prz} is reflected in the fact that $\tau\Big|_S = dz$.
 \item Choose coordinates $(w,\wbar, z, \zbar, \xi)$ on correspondence space such that $\xi\Big|_S = 0$ and $(\xi)^{-1}\Big|_{\iota(S)} = 0$. In a neighbourhood of $S$, the restriction of the contact form to the twistor lines $L_m$ for $m \in M$ is of the form $\tau\Big|_{L_m} = e^{\Phi} d\xi$, defining a function $\Phi$ on correspondence space. Similarly we obtain around $\iota(S)$ a second function $\tilde{\Phi}$ and find the Przanowski function to be $K = - \frac{1}{\Lambda} \left( \Phi\Big|_S + \tilde{\Phi}\Big|_{\iota(S)} \right)$.
\end{itemize}
Not all choices in this procedure are unique, the resulting freedom will be seen to be a gauge freedom of the Przanowski gauge.\\

\subsection{Outline}
This paper is organised as follows. Having introduced the metric of a Quaternion-K\"ahler four-manifold in Przanowski's form \eqref{metric} above, in section \ref{Przfn} we proceed to show that it is indeed anti-self-dual (ASD) and Einstein. Furthermore, we construct a conformally invariant differential operator and consider the associated generalised Laplacian. In section \ref{twistor} we construct the twistor space of a Quaternion-K\"ahler manifold and as a spin-off obtain a Lax Pair for Przanowski's equation. We discuss recursion relations relating solutions to the generalised Laplace equation to cohomology classes on twistor space. At the end of this section, we focus on the linearised Przanowski operator as a special case of the generalised Laplacian and describe deformations of the contact structure on twistor space generated by perturbations of Przanowski's function. In section \ref{topdown} we provide an algorithm to obtain Przanowski's function from twistor data in the double-fibration picture by making a suitable choice of gauge. We then use section \ref{examples} to illustrate this procedure in a few examples: $S^4$, $H^4$, $\C P^2$ and $\widetilde{\C P^2}$, the non-compact version of $\C P^2$ with the Bergmann metric.

\subsection{Notation}
Throughout this paper we will be dealing exclusively with four-dimensional Quaternion-K\"ahler manifolds with non-zero cosmological constant\footnote{In our conventions $R_{ab} = 3 \Lambda g_{ab}$ on an Einstein manifold.} $\Lambda \neq 0$. We will make use of the spinor formalism for four-manifolds, which we introduce in Appendix B. Our conventions are mainly based on \cite{Mason, Dunajski2009}.\\

\section{Differential operators on Quaternion-K\"ahler four-manifolds}
\label{Przfn}
In this section, we will introduce Przanowski's form of a Quaternion-K\"ahler metric on a four-manifold $(M,g)$. Locally one can always find a compatible complex structure on $M$ with complex coordinates $(w, z)$ and complex conjugates $(\wbar, \zbar)$. Of course this need not be true globally, as a counter-example consider $S^4$ which is anti-self-dual Einstein with the round metric, but does not admit a global complex structure. With respect to such a local complex structure the metric can be written in hermitian form as in \eqref{metric}, and the metric is ASD Einstein if and only if $K$ satisfies Przanowski's equation \eqref{Prz} as shown in \cite{Przanowski}. The first part of this assertion will be seen to be true at the end of this section, while the necessity will become clear when recovering Przanowski's formulation from the twistor description of a Quaternion-K\"ahler manifold. We can always find a null tetrad adapted to the complex structure so that $e^{A0'} \in \Lambda^{(1,0)}M$ while $e^{A1'} \in \Lambda^{(0,1)}M$. This reduces the gauge freedom from $SO(4,\C)$ to $GL(2,\C) \cong SL(2,\C) \times \C^{\times}$. Here $SL(2,\C)$ acts on $\spS$ while $\C^{\times}$ is a subgroup of $SL(2,\C)$ acting on $\spS'$ via\footnote{This corresponds to a transformation $o^{A'} \mapsto e^{\Theta} o^{A'}$ and $\rho^{A'} \mapsto e^{-\Theta} \rho^{A'}$ using the notation of Appendix B.} $e^{A0'} \mapsto e^{\Theta} e^{A0'}$ and $e^{A1'} \mapsto e^{-\Theta} e^{A1'}$.  We can fix the gauge freedom by choosing

\begin{align}
\label{nulltetrad}
 e^{00'} = dw, \quad e^{10'} = dz, \quad e^{01'} = -\Kzw d\wbar - \left( \Kzz + \frac{2}{\Lambda} e^{\Lambda K} \right) d\zbar, \quad e^{11'} =  \Kww d\wbar + \Kwz d\zbar  .
\end{align}
Using the abbreviation

\begin{align}
 \label{tildeK}
 \tilde{K} := \Kwz \Kzw - \Kww \left( \Kzz + \frac{2}{\Lambda} e^{\Lambda K} \right),
\end{align}
we obtain for the self-dual two-forms on $M$

\begin{align}
 \label{SDforms}
 \Sigma^{0'0'} = dw \wedge dz, \quad \quad  \Sigma^{0'1'} =  \frac{1}{2} \p \bar{\p} K + \frac{1}{\Lambda} e^{\Lambda K}  dz \wedge d\zbar, \quad \quad \Sigma^{1'1'} = - \tilde{K} d\wbar \wedge d\zbar,
\end{align}
where $\Sigma^{0'1'}$ is the hermitian two-form and $d = \p + \bar{\p}$ is the splitting of the exterior derivative induced by the complex structure. Again, note the Dolbeault types of these forms: $\Sigma^{0'0'} \in \Lambda^{(2,0)}M$, $\Sigma^{0'1'} \in \Lambda^{(1,1)}M$ and $\Sigma^{1'1'} \in \Lambda^{(0,2)}M$. The Hermitian two-form depends only on the choice of complex structure, while the other self-dual two-forms transform with weight $\pm 2$ under the $\C^{\times}$-action, e.g. $\Sigma^{0'0'} \mapsto e^{2 \Theta} \Sigma^{0'0'}$. Using the vector fields

\begin{alignat}{2}
 \nonumber
 \nabla_{00'} &= \p_w, & \quad \quad \quad \quad \nabla_{11'} &= \frac{1}{\tilde{K}} \left[- \left( \Kzz +\frac{2}{\Lambda} e^{\Lambda K} \right) \p_{\wbar} + \Kzw \p_{\zbar} \right], \\
 \label{vf}
 \nabla_{10'} &= \p_z, & \quad \quad \quad \quad \nabla_{01'} &= \frac{1}{\tilde{K}} \left[- \Kwz \p_{\wbar} + \Kww \p_{\zbar} \right],
\end{alignat}
which are dual to the null tetrad, we obtain for the primed connection

\begin{align*}
 \Gamma_{0'0'} &= - \nabla_{A0'} (\ln \Kwbar) e^{A1'}, \quad \quad \quad \quad \Gamma_{1'1'} = \nabla_{A1'} ( \ln \Kw ) e^{A0'},  \\
 \Gamma_{0'1'} &= \frac{1}{2} \left[ \nabla_{A0'} (\ln \tilde{K} - \ln \Kwbar) e^{A0'} + \nabla_{A1'} \left( \ln \Kw \right) e^{A1'}  \right].
\end{align*}
To simplify these expressions we used Przanowski's equation. Our choice of adapted null tetrad \eqref{nulltetrad} leads to a particularly simple form of the primed connection:

\begin{align}
\label{1formID} \Gamma_{0'0'}  \in \Lambda^{(1,0)}M, \quad \Gamma_{1'1'} \in \Lambda^{(0,1)}M, \quad d \Gamma_{0'1'} \in \Lambda^{(1,1)}M, \quad \mbox{with} \quad \Gamma_{0'0'} \wedge d \Gamma_{0'0'} = \Gamma_{1'1'} \wedge d \Gamma_{1'1'} = 0.
\end{align}
At this point we can check directly that the metric \eqref{metric} is ASD and Einstein. To do this, we compute the primed curvature spinor $R_{A'B'}$ and upon substituting \eqref{Prz} find

\begin{align}
\label{Einstein}
R_{A'B'} = \Lambda \Sigma_{A'B'}.
\end{align}
Thus the self-dual Weyl spinor and the trace-free Ricci spinor vanish as claimed. The converse was already shown by Przanowski almost 30 years ago \cite{Przanowski}. We will obtain it from the twistor picture in section \ref{topdown}. Using the exterior derivative of \eqref{Einstein} we can define two one-forms $A$ and $B$ by

\begin{align}
\label{AandB}
 d\Sigma^{0'0'} = \left( B-A \right) \wedge \Sigma^{0'0'}, \quad \quad \quad  d\Sigma^{0'1'} = B \wedge \Sigma^{0'1'}, \quad \quad \quad  d\Sigma^{1'1'} = \left( B+A \right) \wedge \Sigma^{1'1'}.
\end{align}
The Lee form $B$ only depends on the choice of complex structure, while $A$ transforms as $A \mapsto A - 2 d\Theta$ under the $\C^{\times}$-action. In terms of $K$, the forms are given by

\begin{align*}
 A = \p \left( \ln \tilde{K} - 2 \ln \Kwbar \right) + \bar{\p} \left( 2 \ln \Kw \right), \quad \quad \quad \quad \quad
 B = \p \left( \ln \Kw \right) + \bar{\p} \left( \ln \Kwbar \right).
\end{align*}

\textit{Remark:} When $\Kw = \Kwbar$ the Lee form $B$ is exact, so in this case $(M, g)$ is locally conformally K\"ahler. However, $\Kw = \Kwbar$ implies that $(M,g)$ has an isometry \cite{Tod:2006wj, PrzanowskiKilling}. Hence this is an example of the more general correspondence proved in \cite{MDTodconfK} that an ASD Einstein four-manifold is conformally K\"ahler if and only if it has an isometry.\\

More generally, under conformal rescalings where $g \mapsto e^{2 \Omega} g$ the Lee form transforms as $B \mapsto B + 2 d\Omega$, we also have $A \mapsto A + 2 d\Omega$ if we keep $e^{A0'}$ invariant. We can then define a differential operator $D$ acting on sections $f_{l,m}$ of the line bundle $L^{l,m} = \left( \Lambda^{(2,0)}M \right)^{\frac{l}{2}} \otimes \left( \Lambda^{(2,2)}M \right)^{\frac{m}{4}}$ by

\begin{align*}
 D f_{l,m} = \left( d + \frac{l}{2} A - \frac{l+m}{2} B \right) f_{l,m}.
\end{align*}
If the section $f_{l,m}$ transforms as $f_{l,m} \mapsto e^{l \Theta + m \Omega} f_{l,m}$ under the $\C^{\times}$-action and a change of conformal scale, then $D$ is an invariant operator that depends only on the conformal class of the ASD Einstein metric and the choice of compatible complex structure. Taking care to transform the conformal weight of $f_{l,m}$ appropriately under the action of the Hodge star operator we note that $\ast D f_{l,m} \in L^{l,m+2} \otimes \Lambda^3 M$ and we can consider the Laplacian $\ast D \ast D$. When acting on sections of $L^{0,-1}$ it reproduces the conformal Laplacian

\begin{align*}
 \ast D \ast D = \ast d \ast d - \frac{1}{6} R,
\end{align*}
whereas for sections of $L^{0,1}$ we find 

\begin{align*}
 \ast D \ast D = \ast d \ast d + 2 \ast \left( \ast B \wedge d \right) + \frac{3}{4} \ast \left( B \wedge \ast B \right) - \frac{1}{2} \ast d \ast B ,
\end{align*}
the linearised Przanowski operator \eqref{linPrz}. Solutions $\delta K \in \Gamma \left( M, L^{0,1} \right)$ to $\ast D \ast D \; \delta K = 0$ are infinitesimal perturbations of the Przanowski function $K$ and thus correspond to deformations of the underlying Quaternion-K\"ahler manifold.

\section{Twistor theory and Przanowski's function}
\label{twistor}
As is well-known, from any four-dimensional Quaternion-K\"ahler manifold one can construct an associated three-dimensional complex twistor space with a four-parameter family of holomorphic curves called twistor lines and a contact structure \cite{WardSDC, WardWells, Mason, Penrose, Dunajski2009, AHS}. We will first use the more general correspondence for anti-self-dual manifolds to construct the Lax Pair \eqref{LaxPair} for Przanowski's equation as well as a recursion relation relating solutions of a generalised Laplace equation to cohomology classes $H^1 \left( T, \mathcal{O}(k) \right)$. Using the recursion relation, we show how to construct coordinates on $T$ in terms of $K$ and also provide a contour integral for perturbations $\delta K$ of Przanowski's function.\\
We will work with the double-fibration picture, so we need to complexify the underlying manifold $M$. We thus promote $(w,\wbar, z, \zbar)$ to four independent complex variables $(w, \tilde{w}, z, \tilde{z})$ and denote the resulting complex four-manifold by $\M$. From the complex conjugation of the underlying real manifold $M$ we inherit an anti-holomorphic involution

\begin{align*}
 \iota_{\M}: \; \M \longrightarrow \M, \; (w, \tilde{w}, z, \tilde{z}) \longmapsto (\bar{\tilde{w}}, \bar{w}, \bar{\tilde{z}}, \bar{z}). 
\end{align*}
The fixed points of this map allow us to retrieve the real manifold $M$, corresponding to reality conditions\footnote{To make notation more convenient, we use four independent holomorphic coordinates $(w,\wbar, z,\zbar)$ on $\M$, we retrieve $M$ when $(\wbar, \zbar)$ are complex conjugates of $(w,z)$.} $\bar{w} = \tilde{w}$, $\bar{z} = \tilde{z}$.

\subsection{Twistor distribution}
\label{corr}
Consider the primed spin bundle without the zero section, $\F = \spS' \backslash \{ \xi^{A'} = 0 \}$, where $\xi^{A'}$ are coordinates on the fibres of $\spS'$. For every section of $\F$, we obtain a distribution of null two-planes in $T\M$ spanned by two vector fields $ \xi^{A'} \nabla_{AA'}$. Multiplying a spinor section $\xi^{A'}$ by a function on $\M$ leaves the null plane unchanged, to eliminate this redundancy we need to consider the correspondence space $F$ which is obtained by projectivising the fibres of $\F$. These are then no longer copies of $\C^2$ without the origin but $\C P^1$s. The space $\F$ can be understood as a holomorphic line bundle over $F$, the points in the fibre representing different multiples of a given null plane. When restricted to a fibre of $F$ over $\M$, this line bundle is just the tautological bundle $\C^2 \longrightarrow \C P^1$.\\
Note that the action of the involution $\iota_{\M}$ can be extended to an involution $\iota$ of $F$. $\iota_{\M}^*$ pulls back $(2,0)$-forms to $(0,2)$-forms and therefore $\iota$, while leaving the fibres over real points in $\M$ invariant, acts as the antipodal map on such a fibre.\\
Parallel transport with respect to the Levi-Civita connection maps null planes to null planes, giving rise to a homogeneous one-form $\tau$ of degree two on $\F$,

\begin{align}
 \label{contactform}
 \tau = \xi_{A'} \left( d\xi^{A'} + \xi^{B'} {\Gamma^{A'}}_{B'} \right).
\end{align}
Using $\tau$, we can lift the vector fields $\xi^{A'} \nabla_{AA'}$ to $\F$ to obtain\footnote{Here $AA'$ are one-form indices, so $\Gamma_{B'C'} = \Gamma_{AA'B'C'} e^{AA'}$.}

\begin{align}
 \label{dsubA}
 d_A = \xi^{A'} \nabla_{AA'} - \xi^{A'} \xi^{B'} {\Gamma_{AA'B'}}^{C'} \p_{\xi^{C'}}.
\end{align}
Since the Euler vector field

\begin{align*}
 \Upsilon &= \xi^{0'} \p_{\xi^{0'}} + \xi^{1'} \p_{\xi^{1'}}
\end{align*}
lies in the kernel of the contact form $\tau$, the vector fields $d_A$ are only determined up to the addition of terms proportional to $\Upsilon$. None the less the choice given in \eqref{dsubA} is a canonical one starting from the metric in Przanowski's form with its compatible complex structure and the null tetrad \eqref{nulltetrad}. Since by definition $d_A \hook \tau = 0$, the vector fields $d_A$ form a distribution on $\F$ that lies within the kernel of $\tau$, called the twistor distribution. It is well-known \cite{Penrose} that this twistor distribution is integrable if and only if $(\M,g)$ is ASD. In general the following identity

\begin{align}
 \label{d0d1}
 \left[ d_0, d_1 \right] = \xi^{A'} {\Gamma_{AA'}}^{AB} d_B
\end{align}
holds for manifolds with vanishing self-dual Weyl spinor.

\subsection{Lax Pair}
\label{lax}
While the integrability of the twistor distribution $<d_0, d_1>$ is equivalent to the anti-self-duality of $(\M,g)$, the fact that $K$ satisfies Przanowski's equation \eqref{Prz} is sufficient but not strictly necessary for this. To obtain a Lax Pair consider the modified vector fields

\begin{align*}
 \tilde{\nabla}_{A0'} = \nabla_{A0'}, \quad \quad \quad
 \tilde{\nabla}_{01'} = - \frac{\tilde{K} }{e^{\Lambda K} \Kw \Kwbar} \; \nabla_{01'} , \quad \quad \quad \tilde{\nabla}_{11'} = - \frac{ \tilde{K} }{e^{\Lambda K} \Kw \Kwbar} \; \nabla_{11'},
\end{align*}
which reduce to \eqref{vf} if and only if \eqref{Prz} is satisfied. Similarly\footnote{The symbol $\hook$ denotes contraction.}

\begin{align*}
\tilde{d}_A = \xi^{A'} \nabla_{AA'} - \xi^{A'} \left( \xi^{B'} \tilde{\nabla}_{AB'}  \hook {\Gamma^{C'}}_{A'} \right) \p_{\xi^{C'}},
\end{align*}
reduces to $d_A$ if and only if \eqref{Prz} holds. We now introduce a trivialisation of $\F$ over $F$ based on the standard trivialisation of the tautological line bundle over $\C P^1$: Consider

\begin{align}
\label{trivF}
 U_0 = \left\{ x \in \F \; \Big| \; \xi^{0'} \neq 0 \right\}, \quad \quad U_1 = \left\{ x \in \F \; \Big| \; \xi^{1'} \neq 0 \right\}.
\end{align}
On $U_0$, define $\xi = \frac{ \xi^{1'} }{ \xi^{0'} }$ and on $U_1$, define\footnote{In most of this paper when considering $\F$, we will work over $U_0$. The formulae valid over $U_1$ can usually be easily inferred.} $\eta = \frac{ \xi^{0'}}{\xi^{1'}}$. Holomorphic functions on $U_0$ homogeneous of degree zero, which can be regarded as functions on $F$ holomorphic in $\xi$, are annihilated by the Euler vector field $\Upsilon$ and thus $\p_{\xi^{A'}}$ acts as $\frac{\xi_{A'}}{(\xi^{0'})^2} \p_{\xi}$. So the projection of $\tilde{d}_A$ to $TF$ is

\begin{align*}
 \tilde{l}_A = \xi^{A'} \nabla_{AA'} + (\xi^{0'})^{-2} \xi^{A'} \xi^{B'} \xi^{C'} \left( \tilde{\nabla}_{AA'} \hook \Gamma_{B'C'} \right) \p_{\xi},
\end{align*}
or in terms of $K$ we find in inhomogeneous form the vector fields $l_A$ as in \eqref{LaxPair}. We show in Appendix A that $[l_0, l_1 ] = 0$ if and only if Przanowski's equation \eqref{Prz} is satisfied, so \eqref{LaxPair} are a Lax Pair for Przanowski's equation, showing that it is an integrable equation.

\subsection{Recursion relations}
\label{twistorT}
Returning to the construction of the twistor space we recall that we have an integrable distribution $<d_0, d_1>$ in $\F$, so we can consider the quotient space $\T = \F \Big\slash <d_0, d_1>$, a four-dimensional complex manifold. The vector fields $d_A$ project to non-zero vector fields $l_A$ on $F$, so we can also consider $T = F \Big\slash <l_0, l_1>$, a three-dimensional complex manifold. Now a point $p \in T$ corresponds to an integral surface $\Sigma$ of the twistor distribution in $F$. We can restrict the line bundle $\F$ to this integral surface to obtain a line bundle $\F_{\Sigma}$. However this line bundle has to be trivial, since we can find a global trivialisation over $\Sigma$ using the leaves of the distribution $<d_0,d_1>$. Thus by construction $\T$ is a line bundle over the twistor space $T$ and if we pull back $\T$ to the correspondence space $F$ we recover $\F$.\\
The fibres of $\F$ over $\M$ become a four-parameter family of copies of $\C^2 \backslash \{ 0 \}$ in $\T$, regarded as $\C P^1$-fibres of $F$ we obtain the twistor lines in $T$. Again, $\T$ restricted to such a twistor line is the standard bundle over $\C P^1$. Furthermore, the involution $\iota$ descends naturally from $F$ to $T$.\\
We now want to explain how one can construct solutions to generalised Laplace equations from elements of $H^1 (\T, \mathcal{O}(k))$, or conversely construct those cohomology classes from solutions of the generalised Laplace equation using a recursion relation. So suppose $\varPsi \in H^1 (\T, \mathcal{O}(k))$, we can pull $\varPsi$ back to $\F$ to obtain $\Psi \in H^1 (\F, \mathcal{O}(k))$ satisfying

\begin{align}
 \label{dAPsi}
 d_A \Psi = 0.
\end{align}
Since $\Psi$ is an element of the first cohomology group, its domain can be assumed to be $U_0 \cap U_1$ where $\xi^{0'} \neq 0$, so we can trivialise\footnote{Different trivialisations will only shift the parameter $n$ in the ensuing discussion.} and write $\Psi = \sum\limits_n \left( \xi^{0'} \right)^{(k-n)} \left(\xi^{1'}\right)^n \psi_n$ where $\psi_n$ are functions on $M$. But then

\begin{align*}
 d_A \Psi= \sum\limits_{n = -\infty}^{\infty} \left( \xi^{0'} \right)^{(k-n)} \left(\xi^{1'}\right)^n \xi^{A'} \left( \nabla_{AA'} + \left(n-\frac{k}{2} \right) A_{AA'} - \frac{k}{4} B_{AA'} \right) \psi_n,
\end{align*}
where the bracket on the right side of the equation defines\footnote{The one-forms $A$ and $B$ were defined in \eqref{AandB}.} a linear first-order operator $\tilde{d}^{(n,k)}_{AA'}$ on $M$,

\begin{align*}
\tilde{d}^{(n,k)}_{AA'} =  \nabla_{AA'} + \left(n-\frac{k}{2} \right) A_{AA'} - \frac{k}{4} B_{AA'}.
\end{align*}
From \eqref{dAPsi} we obtain the recursion relations

\begin{align}
 \label{RecRel}
 \tilde{d}^{(n+1,k)}_{A0'} \psi_{n+1} = - \tilde{d}^{(n,k)}_{A1'} \psi_{n}.
\end{align}
For \eqref{RecRel} to be consistent, $\psi_{n}$ has to satisfy an integrability condition. Since \eqref{d0d1} implies

\begin{align}
 \label{WZ0}
\epsilon^{AB} \xi^{A'} \xi^{B'} \left[ \tilde{d}^{(n,k)}_{AA'} \tilde{d}^{(n,k)}_{BB'} - \left( {\Gamma_{BA'B'}}^{C'} + {\Gamma_{CB'B}}^C {\epsilon_{A'}}^{C'} \right) d^{(n,k)}_{AC'} \right] = 0,
\end{align}
cross-differentiating \eqref{RecRel} and imposing \eqref{WZ0} requires

\begin{align*}
 \left( \tilde{d}^{(n,k)}_{AA'} + {\Gamma_{AB'A'}}^{B'} + {\Gamma_{BA'A}}^B + B_{AA'} \right) \tilde{d}^{(n,k) \; AA'} \psi_{n} = 0.
\end{align*}
Rewriting this expression covariantly, we find

\begin{align}
 \label{intcond}
 \ast D \ast D \; \psi_{n} = 0
\end{align}
if we assign $\psi_{n}$ weight $(l,m) = (k-2n, 2n-\frac{k}{2})$. So starting with $\psi_{n}$ satisfying this integrability condition we can use the recursion relations \eqref{RecRel} to determine $\psi_{n+1}$, where $\ast D \ast D \; \psi_{n+1} = 0$ is automatically guaranteed, again using \eqref{WZ0}. This allows us to use the recursion relations to define $\psi_{n+2}$, and so forth. Thus a single coefficient $\psi_n$ satisfying \eqref{intcond} is sufficient to determine $\varPsi \in H^1 \left(T, \mathcal{O}(k) \right)$. Conversely, given such a $\varPsi$, each coefficient will satisfy a second-order integrability condition.\\

\textit{Remark:} We only needed the fact that $M$ is Einstein to define $A$ and $B$ in \eqref{AandB}. If instead we define
\begin{align*}
 A = \Gamma_{A \left( A'0'1' \right)} e^{AA'}, \quad \quad \quad \quad B = -2 \Gamma_{A1'0'0'} e^{A0'} + 2 \Gamma_{A0'1'1'} e^{A1'},
\end{align*}
then the recursion relations \eqref{RecRel} and the integrability condition \eqref{intcond} make sense on any anti-self-dual four-manifold. On an Einstein manifold this definition for $A$ and $B$ is equivalent with \eqref{AandB}.\\

We will now show that one can use the correspondence above to construct coordinates on $T$ from solutions to \eqref{intcond}. Consider solutions of \eqref{intcond} with weight $(0,0)$, one can check that holomorphic functions $f(w,z)$ on $M$ as well as anti-holomorphic functions $f(\wbar, \zbar)$ on $M$ are examples. If we choose $\psi_0$ to be an anti-holomorphic function on $M$, then the recursion relations \eqref{RecRel} imply that $\psi_{n} = 0$ for all negative coefficients and so $\Psi$ will in fact be an element of $H^0 \left( U_0, \mathcal{O}(0) \right)$ that descends to $T$. Therefore we can recursively construct coordinates of $T$ on the image of $U_0$ under the canonical projection to $T$ by setting $\psi_0$ equal to $\wbar$, $\zbar$ or a constant. Similarly, on the image of $U_1$ we can start with $\psi_0$ equal to $w$, $z$ or a constant.\\

\subsection{Perturbations}
\label{contact}
The twistor space with its twistor lines only encodes the conformal structure of $\M$, the information necessary to retrieve an Einstein metric within the conformal class is contained in a contact structure on $T$. Essentially all that is needed to fix the metric within the conformal class is a scale, which is specified uniquely by the symplectic structures $\epsilon_{AB}$ and $\epsilon_{A'B'}$ of $\spS$ and $\spS'$. Since the basis $\xi^{A'}$ of $\spS'$ is normalised such that $\epsilon_{0'1'} \xi^{0'} \xi^{1'} = 1$ and similarly $\epsilon_{AB}$ is contained within the definition of $\Gamma_{A'B'}$, all this information is stored in the one-form $\tau$ on $\F$, given by \eqref{contactform}. It corresponds to a one-form $\tau_F$ quadratic in $\xi$ on $F$, where

\begin{align*}
 \tau_F = d\xi - \Gamma_{0'0'} - 2 \xi \Gamma_{0'1'} - \xi^2 \Gamma_{1'1'}.
\end{align*}
By construction, $d_A \hook \tau = 0$ and the Einstein equations imply $\mathcal{L}_{d_A} \tau = 0$. Therefore $\tau$ descends to a holomorphic one-form homogeneous of degree two on $\T$, where it satisfies $d \tau \wedge d\tau \neq 0$ and so defines a holomorphic symplectic structure on $\T$. This corresponds to a holomorphic contact structure on $T$, determined by a contact one-form $\tau_T$, the relationship between the symplectic structure on $\T$ and the contact structure on $T$ as well as their deformations are illuminated in \cite{LeBrun}. According to Darboux' Theorem, one can always choose canonical coordinates on $T$ such that $\tau_T = dx - y dt$. Furthermore, the pull-back of $\tau_T$ to $F$ is proportional to $\tau_F$.\\
Now recall that for $(l,m)=(0,1)$ or equivalently $(n,k)=(1,2)$ equation \eqref{intcond} is the linearised Przanowski equation. Thus by definition the coefficient $\psi_1$ of every element $\varPsi \in H(T, \mathcal{O}(2))$ is a solution $\delta K \in L^{0,1}$ of \eqref{intcond}. Indeed perturbations of Quaternion-K\"ahler metrics are known to be generated by elements of $H^1 \left( T, \mathcal{O}(2) \right)$ \cite{LeBrun}. To see this, regard a representative $\varPsi$ of this cohomology class as a Hamiltonian of a one-parameter family of symplectic transformations. So $d \varPsi = d \tau \left( \theta, \cdot \right)$ where $\theta \in H^1 \left( T, \Theta(0) \right)$ is an element of the first cohomology group with values in the sheaf of holomorphic vector fields. Therefore $\theta$ encodes a deformation of the holomorphic symplectic structure of $\T$ and consequently a deformation of the metric of $M$. For details on complex deformations see \cite{Kodaira}. We can obtain $\delta K$ from $\Psi \in H^1 \left( F, \mathcal{O}(2) \right)$, Cauchy's integral formula

\begin{align*}
 \delta K &= \frac{1}{2 \pi i} \oint\limits_{\Gamma} \frac{\psi_1}{\xi} d\xi 
          = \frac{1}{2 \pi i} \oint\limits_{\Gamma} \frac{ \varPsi \xi_{A'} d \xi^{A'} }{\left( \xi^{A'} o_{A'} \right)^2 \left( \xi^{A'} \rho_{A'} \right)^2}
\end{align*}
reproduces \eqref{contint}. Here the constant spinors $o_{A'} = (1,0)$ and $\rho_{A'} = (0,1)$ are determined by the choice of complex structure on $M$ and $\Gamma$ is any contour around the equator of $\C P^1$.\\
This is similar to the work of \cite{MDMason} but for non-zero cosmological constant, and extends results of \cite{Neitzke:2007ke, Alexandrov:2009vj} to Quaternion-K\"ahler four-manifolds with no isometries.

\section{Przanowski's function from Twistor data}
\label{topdown}
We now want to explain how to derive the existence of Przanowski's function as well as the second-order partial differential equation \eqref{Prz} it satisfies from the description of a Quaternion-K\"{a}hler four-manifold by its twistor space. From this, we will obtain an algorithm to extract Przanowski's function and a compatible complex structure from twistor data. This is similar to the procedure in \cite{Alexandrov:2009vj}, however adapted to the double-fibration picture and we don't require any information about the metric or K\"ahler structure on twistor space.\\
We thus assume we are given a three-dimensional complex manifold $T$ with a four-parameter family $\M$ of twistor lines. Furthermore we have a holomorphic contact structure determined by a homogeneous one-form $\tau$ of degree two on $\T$, such that $\tau_T(Q) \neq 0$ for any nonzero vector $Q$ tangent to one of the twistor lines. Let $R = 12 \Lambda$ be the scalar curvature and cosmological constant of the Quaternion-K\"{a}hler manifold respectively, the real structure of the underlying manifold $M$ is encoded in an involution $\iota$ on $T$.\\
Using this information, we will show that locally one can always choose an integrable complex structure on $M$ with coordinates $(w,z,\wbar, \zbar)$ such that the adapted null tetrad with $e^{00'}= dw$, $e^{10'}= dz$ and $e^{A1'} \; \in \; \Lambda^{(0,1)}M$ leads to primed connection one-forms of the following form:

\begin{align}
\label{Przconn}
\Gamma_{0'0'} = f d\zbar, \quad \quad \quad \quad \Gamma_{1'1'} = g dz, \quad \quad \quad \quad  d\Gamma_{0'1'} = \frac{\Lambda}{2} \bar{\p} \p K,
\end{align}
for some complex-valued functions $f$, $g$ and $K$ on $\M$ with $K = \frac{1}{\Lambda} \ln fg$. Under the induced real structure on $M$, $K$ is a real function and we will identify it with Przanowski's function.\\

\textit{Remark 1:} In this procedure the coordinates $(w,z,\wbar, \zbar)$ are determined only up to holomorphic coordinate transformations given by

\begin{align}
 \label{coordtrafo}
 w \rightarrow w'(w,z), \;\; z \rightarrow z'(z), \;\; \wbar \rightarrow \wbar'(\wbar,\zbar), \;\; \zbar \rightarrow \zbar'(\zbar).
\end{align}
Under such a change of coordinates, $K$ will transform as

\begin{align*}
 K(w,z,\wbar, \zbar) \rightarrow K(w', z', \wbar', \zbar') - \frac{1}{\Lambda}\ln (\p_z z') - \frac{1}{\Lambda} \ln (\p_{\zbar} \bar{z}').
\end{align*}
One can check that this gives rise to the same metric.\\

\textit{Remark 2:} The Przanowski function $K$ determines $d\Gamma_{0'1'}$, a non-degenerate closed two-form. This symplectic form, which is neither compatible with the metric nor covariant, admits both $(w,z, -K_w, -K_z)$ as well as $(\wbar, \zbar, K_{\wbar}, K_{\zbar})$ as canonical coordinates,

\begin{align*}
 d\Gamma_{0'1'} = d K_{\wbar} \wedge d\wbar + d K_{\zbar} \wedge d\zbar = - d K_w \wedge dw - d K_z \wedge dz.
\end{align*}
Thus $K(w,z,\wbar, \zbar)$ can be regarded as the generating function for the symplectic transformation that maps ``initial positions'' $(w,z)$ to ``final positions'' $(\wbar, \zbar)$. This is a remnant of the interpretation of the heavenly function as a transition function on Hyper-K\"ahler manifolds \cite{MasonNewman}.\\

We now give the details of the construction. Suppose that $x,y,t$ are local holomorphic coordinates on $T$ and $\tau_T = dx - y dt$. To obtain a local complex structure, choose a holomorphic surface $S'_1$ in the twistor space $T$ transversal to the twistor lines. This may not be possible for all lines, the complex structure is not defined for points in $M$ whose twistor lines are tangent to $S'_1$, we may wish to exclude these points from $M$. For instance we can choose $S'_1 = \{ p \in T \; | \; t = 0 \}$. The pre-image of this surface in the correspondence space $F$ is a four-dimensional holomorphic surface $S_1$ which is also a section $s$ of the $\C P^1$-bundle over the base manifold $\M$. We can use $x ,y$ as coordinates on $S'_1$, pulled back to $F$ the one-forms $dx$ and $dy$ annihilate the twistor distribution. We define $\wbar = x\Big|_{S_1}$, $\zbar = y\Big|_{S_1}$ and

\begin{align*}
 \Lambda^{(0,1)}M = < d\wbar, d \zbar >.
\end{align*}
From $S'_2 = \iota ( S'_1 )$ we obtain two more coordinates $w = \iota(x)\Big|_{S_2}$ and $z = \iota(y)\Big|_{S_2}$. By construction these will be complex conjugates of $(w,z)$ on the underlying real manifold $M$. We use them to define $\Lambda^{(1,0)} M = <dw, dz>$. As any two different null planes through a point span the entire tangent space, the functions $(w,z, \wbar, \zbar)$ will be independent on $M$. Locally this defines an integrable complex structure compatible with the metric \cite{Salamon}.\\
Since a contact structure on a three-dimensional manifold has no integral submanifolds of dimension higher than one, the one-form $\tau_T$ is non-zero when restricted to any two-dimensional surface. Darboux' Theorem ensures that we can always choose our coordinates on $M$ such that $\tau_T = f' d\zbar$ on $S_1$ and $\tau_T = g' dz$ on $S_2$. This step is well-defined up to transformations of the form \eqref{coordtrafo} and is the origin of the asymmetry between $w$ and $z$ in the Przanowski equation.\\
By construction the metric $g$ of $M$ is Hermitian with respect to the coordinates $(w, z, \wbar, \zbar)$, choosing an adapted tetrad with $e^{A0'} \in \Lambda^{(1,0)}M$ and $e^{A1'} \in \Lambda^{(0,1)}M$ reduces the gauge freedom to $GL(2,\C)$. In the trivialisation \eqref{trivF} The pre-images of the hypersurfaces $S'_1$ and $S'_2$ are then given by $S_1 = \{ p \in F \; | \; \xi = 0 \}$ and $S_2 = \{ p \in F \; | \; \eta = 0 \}$. For convenience we choose $e^{00'} = dw$, $e^{10'} = dz$ to fix the frame uniquely, to find $K$ we need the explicit expressions of $e^{A1'}$. They can be obtained from $2 \Sigma^{0'1'}$, which is the hermitian two-form compatible with the metric $g$ and the complex structure, using the primed connection one-forms.\\
First, to show that the primed connection is always of the form \eqref{1formID} in a frame adapted to the complex structure, we start by classifying $\Gamma_{A'B'}$ according to their Dolbeault-type. By definition,

\begin{align*}
de^{AA'} = ( {\Gamma^{AA'}}_{CC'} )_{BB'} e^{BB'} \wedge e^{CC'}.
\end{align*}
Now integrability of the complex structure means that

\begin{alignat*}{2}
  {\Gamma^{A1'}}_{[00'10']} &= 0, \quad \quad \quad \quad & {\Gamma^{A0'}}_{[01'11']} &= 0.
\end{alignat*}
Therefore, using anti-symmetry in the Lie algebra indices and considering separately the cases $A=0$ and $A=1$,

\begin{align*}
 \Gamma_{0'0'} \in \Lambda^{(0,1)}, \quad \quad \quad  \Gamma_{1'1'} \in \Gamma^{(1,0)}.
\end{align*}
Now recall the components of the ASD Einstein equation,

\begin{align}
\label{QKID}
 d \Gamma_{0'0'} + 2 \Gamma_{0'0'} \wedge \Gamma_{0'1'} = \Lambda \Sigma_{0'0'}, \quad d\Gamma_{0'1'} + \Gamma_{0'0'} \wedge \Gamma_{1'1'} = \Lambda \Sigma_{0'1'}, \quad d\Gamma_{1'1'} + 2 \Gamma_{0'1'} \wedge \Gamma_{1'1'} = \Lambda \Sigma_{1'1'}.
\end{align}
Denoting the component of $\Gamma_{0'1'}$ in $\Lambda^{(a,b)}M$ by $\Gamma^{(a,b)}_{0'1'}$, the first of equations \eqref{QKID} splits up into

\begin{align*}
 d\Gamma^{(0,2)}_{0'0'} + 2 \Gamma_{0'0'} \wedge \Gamma^{(0,1)}_{0'1'} = \Lambda \Sigma_{0'0'},  \quad \quad \quad \quad \quad d\Gamma^{(1,1)}_{0'0'} + 2 \Gamma_{0'0'} \wedge \Gamma^{(1,0)}_{0'1'} = 0.\\
\end{align*}
Therefore $d\Gamma^{(0,2)}_{0'0'} \wedge \Gamma_{0'0'} = d\Gamma^{(1,1)}_{0'0'} \wedge \Gamma_{0'0'} = 0$, and thus
\begin{align*}
 d\Gamma_{0'0'} \wedge \Gamma_{0'0'} = 0.
\end{align*}
Similarly, the remaining identities in \eqref{1formID} follow. Now recall that on $S'_1$ we have $\tau_T = d\zbar$ and on $S'_2$ we find $\tau_T = dz$. Similarly $\tau_F = - \Gamma_{0'0'}$ on $S_1$ and $\tau_F = - \Gamma_{1'1'}$ on $S_2$. But the contact structure on $T$ is induced from the one on $F$, so the pull-back of $\tau_T$ to $F$ is proportional to $\tau_F$, consequently

\begin{align*}
 \Gamma_{0'0'} = f d\zbar, \quad \quad \quad \quad \quad \Gamma_{1'1'} = g dz,
\end{align*}
for some complex-valued functions $f$ and $g$. Furthermore, since $d\Gamma_{0'1'}$ is a closed (1,1)-form, it can be written as

\begin{align}
\label{ddbar}
 d\Gamma_{0'1'} = \frac{\Lambda}{2} \bar{\p} \p K
\end{align}
for some complex-valued function $K$. So far we have established \eqref{Przconn}, it remains to show that $K$ is indeed the Przanowski function and real.\\
From \eqref{ddbar} $K$ is determined only up to the addition of two functions $c(w,z)$ and $\tilde{c}(\wbar,\zbar)$. Using equations \eqref{QKID} it is easy to show that one can choose $c$ and $\tilde{c}$ such that

\begin{align}
 \label{fg}
 \Gamma_{0'0'} = e^{\Lambda K} K_w d\zbar, \quad \quad \quad \quad \Gamma_{1'1'} = \frac{1}{K_w} dz, \quad \quad \quad \quad \Gamma_{0'1'} = \frac{1}{2} \left( d(\ln(K_w)) + \Lambda \p K \right),
\end{align}
together with the self-dual two-forms $\Sigma^{A'B'}$ as in \eqref{SDforms}. Then $\Sigma^{0'0'} \wedge \Sigma^{1'1'} = -2 \Sigma^{0'1'} \wedge \Sigma^{0'1'}$, which follows from \eqref{SigmaID}, is equivalent to Przanowski's equation \eqref{Prz}. We saw earlier that $2 \Sigma^{0'1'}$ is the hermitian two-form with respect to the complex structure and metric $g$ on $M$, so $g$ must be given by \eqref{metric}. Thus $K$ in \eqref{ddbar} is indeed Przanowski's function and real. To determine $K$ explicitly, observe that \eqref{fg} implies

\begin{align}
\label{K}
 K = \frac{1}{\Lambda} \ln {fg}.
\end{align}
Evaluating the restriction of $\tau$ to the sections $(\xi^{0'}, \xi^{1'}) = (1,0)$ and $(0,1)$ of $F$ provides $f$ and $g$ and thus yields Przanowski's function using \eqref{K}.

\section{ Examples}
\label{examples}
We will demonstrate the procedure of writing a Quaternion-K\"ahler metric in Przanowski's form explicitly for a few examples: $S^4$, $H^4$, $\C P^2$ with the Fubini-Study metric and $\widetilde{\C P^2}$ with the Bergmann metric. The first two cases are conformally flat with negative and positive scalar curvature respectively\footnote{According to our conventions $S^4$ has negative scalar curvature, following \cite{WardSDC, Dunajski2009}.} and are treated in \cite{Alexandrov:2009vj}. The other two examples instead are non-trivial, the Fubini-Study metric has negative scalar curvature and the Bergmann metric positive scalar curvature. The twistor data for the Fubini-Study metric is given in \cite{AHS, WardSDC} and can be easily adapted to accommodate for the Bergmann metric\footnote{See \cite{Alexandrov:2009vj} for a description of the latter twistor space with Przanowski's function in a different gauge.}.

\subsection{$S^4$ and $H^4$}
$S^4$ and $H^4$ are conformally flat, the only difference in their twistor data arises in the contact structure. However, it is convenient to use slightly different parametrisations of the twistor lines. Defining $\epsilon$ to be the sign of the cosmological constant, $\Lambda = \epsilon |\Lambda|$, we can treat both cases simultaneously by including $\epsilon$ as a parameter. We will initially normalise $\Lambda$ to $\pm 1$ and return to the general case at the end. To obtain $S^4$, set $\epsilon = -1$, to obtain $H^4$, set $\epsilon = 1$. The twistor space is $\C P^3$ for $S^4$ and an open subset thereof for $H^4$. Parametrising $\C P^3$ by homogeneous coordinates $(u_0, u_1, v_0, v_1)$, the twistor lines are given by

\begin{alignat*}{2}
 u_0 &= \frac{\xi^{0'}}{\sqrt{1-\epsilon|w|^2(1+|z|^2)}}, \quad \quad \quad & \quad \quad \quad v_0 &= \frac{w \xi^{0'} +  \wbar \zbar \xi^{1'}}{\sqrt{1-\epsilon|w|^2(1+|z|^2)}},\\
 u_1 &= \frac{\xi^{1'}}{\sqrt{1-\epsilon|w|^2(1+|z|^2)}}, \quad \quad \quad & \quad \quad \quad v_1 &= \frac{wz \xi^{0'} - \wbar \xi^{1'}}{\sqrt{1-\epsilon|w|^2(1+|z|^2)}}.
\end{alignat*}
Here $(w, z, \wbar, \zbar)$ are coordinates on $\M$, the four-parameter family of twistor lines, and $(\xi^{0'}, \xi^{1'})$ are homogeneous coordinates along such a line. The twistor lines are invariant under the involution $\iota$ if $(\wbar, \zbar)$ are complex conjugates of $(w,z)$. We specify a contact structure by

\begin{align*}
 \tau = \varepsilon^{AB} \left( u_A d u_B + \epsilon v_A d v_B \right).
\end{align*}
The parametrisation of the twistor lines is chosen so that when restricted to a line the contact form is $\tau\Big|_{L_m} = \varepsilon_{A'B'} \xi^{A'} d \xi^{B'}$ so $(\xi^{0'}, \xi^{1'})$ is a normalised basis of $\spS'$. On $U_0=\{ u \in \C P^3 \; | \; u_0 \neq 0 \}$ we can introduce inhomogeneous coordinates

\begin{align*}
\left( \frac{u_1}{u_0}, \frac{v_0}{u_0}, \frac{v_1}{u_0} \right) = \left( \xi, w +  \wbar \zbar \xi, wz - \wbar \xi \right), 
\end{align*}
and choose a holomorphic surface in $T$ by setting $S_1 = \{p \in T \; | \; \xi = 0 \}$. On $U_1 = \{ u \in \C P^3 \; | \; u_1 \neq 0 \}$ we use coordinates 

\begin{align*}
\left( \frac{u_0}{u_1}, \frac{v_0}{u_1}, \frac{v_1}{u_1} \right) = \left( \eta, \wbar \zbar + \eta w, - \wbar + \eta wz \right), 
\end{align*}
and find $S_2 = \iota ( S_1 ) = \{p \in T \; | \; \eta = 0 \}$. This yields a complex structure on $M$ with holomorphic coordinates $w,z$ induced from

\begin{align*}
 w = \frac{v_0}{u_0}\Bigg|_{S_1}, \quad \quad \quad \quad wz = \frac{v_1}{u_0}\Bigg|_{S_1}.
\end{align*}
As complex conjugates we obtain $\wbar, \zbar$ from $S_2$. Now observe that

\begin{align*}
 \tau\Bigg|_{S_1} = \frac{ \epsilon w^2 dz}{1-\epsilon|w|^2(1+|z|^2)}, \quad \quad \quad \quad \quad \tau\Bigg|_{S_2} = \frac{ \epsilon \wbar^2 d\zbar}{1-\epsilon|w|^2(1+|z|^2)},
\end{align*}
so we can use \eqref{K} to find Przanowski's function,

\begin{align*}
 K =  \frac{2}{\Lambda} \ln \left[ \frac{|w|^2}{1-\epsilon|w|^2(1+|z|^2)} \right],
\end{align*}
where we have now included the cosmological constant as a free parameter.

\subsection{$\C P^2$ and $\widetilde{\C P^2}$}
As a non-trivial example we now consider $\C P^2=SU(3) \Big\slash U(2)$ with the Fubini-Study metric, which has negative scalar curvature, and its non-compact version $\widetilde{\C P^2} = SU(2,1) \Big\slash U(2)$ with the Bergmann metric, which has positive scalar curvature. Recall that $\C P^2$ is the space of lines through the origin in $\C^3$, the Fubini-Study metric is induced from a Hermitian form with signature $(+++)$. In constrast, for $\widetilde{\C P^2}$ consider $\C^3$ equipped with a Hermitian form with signature $(++-)$. Then $\widetilde{\C P^2}$ is the space of timelike lines and the Hermitian form induces the Bergmann metric. Although not conformally equivalent, we can again treat both cases simultaneously by introducing a parameter $\epsilon$ where $\Lambda = \epsilon |\Lambda|$, alternatively $\epsilon$ is the negative of the third eigenvalue of the Hermitian form. We initially assume $\Lambda = \pm 1$. The twistor space $T$ is the flag manifold of $\C^3$, so every point of $T$ consists of a pair $(l,p)$ where $p$ is a plane in $\C^3$ and $l$ is a line in $p$, both containing the origin. For $\widetilde{\C P^2}$ we furthermore require that $l$ be space-like and that $p$ contain a time-like direction. Using homogeneous coordinates, we can write any point in $T$ as a pair $(l^j, p_j)$ where $j = 0,1,2$ and $p_j l^j  = 0$.\\
Next we need the twistor lines, these are of the following form: let $P$ be a plane in $\C^3$ and $L$ a line in $\C^3$ \textit{not} in $P$. For $\widetilde{\C P^2}$ we need $L$ to be time-like while $P$ must be spanned by two space-like vectors. Then a twistor line in $T$ is given by all pairs $(l,p)$ where $p$ contains $L$ and where the two planes $p$ and $P$ intersect in $l$. Using homogeneous coordinates for $P$ and $L$, the equation for the twistor line is $P_j l^j = p_j l^j = p_j L^j = 0$. If we write $P_j = (W, Z, 1)$ and $L^j = (\tilde{W}, \tilde{Z}, 1)$ we can use $(W, Z, \tilde{W}, \tilde{Z})$ as coordinates\footnote{These serve as coordinates on all of $\widetilde{\C P^2}$, but only on a coordinate patch of $\C P^2$.} on $\M$. One can check that

\begin{align}
\label{CP2TL}  l^j &= \left( - (1+Z\tilde{Z}) \xi^{1'}, \frac{\xi^{0'}}{1+Z\tilde{Z}}+W\tilde{Z} \xi^{1'}, \frac{-Z\xi^{0'}}{1+Z\tilde{Z}}+W \xi^{1'} \right),\\
\nonumber  p_j &= \left( \frac{\xi^{0'}}{1+W\tilde{W}+Z\tilde{Z}} , \frac{-\tilde{W}Z \xi^{0'}}{(1+Z\tilde{Z})(1+W\tilde{W}+Z\tilde{Z})} +\xi^{1'}, \frac{-\tilde{W}\xi^{0'}}{(1+Z\tilde{Z})(1+W\tilde{W}+Z\tilde{Z})} -\tilde{Z}\xi^{1'} \right),
\end{align}
satisfy the defining equations of a twistor line. To fix a metric within the conformal structure, we chose a contact form

\begin{align*}
 \tau = \frac{1}{2} \left( p_j d l^j - l^j d p_j \right).
\end{align*}
The parametrisation \eqref{CP2TL} of the twistor lines has been chosen to ensure that the restriction of the contact form to the twistor lines is in canonical form,

\begin{align*}
 \tau\Bigg|_{L_m} = \xi^{0'} d \xi^{1'} - \xi^{1'} d \xi^{0'}.
\end{align*}
A further difference between the Fubini-Study metric and the Bergmann metric on the level of their twistor description arises when we describe the involution on $T$. This involution $\iota$ is induced from the Hermitian form on $\C^3$ which defines an anti-linear map from $\C^3$ to the dual space, and thus an anti-holomorphic map from $T$ to itself. Under this map $\iota$ a pair $(l,p)$ is mapped to $(\bar{p}, \bar{l})$, pairs invariant under this map correspond to real twistor lines. Applied to a twistor line $(L^j, P_j)$ we obtain the reality conditions

\begin{align*}
 \bar{W} = - \epsilon \tilde{W}, \quad \quad \quad \bar{Z} = - \epsilon \tilde{Z}.
\end{align*}
For the Bergmann metric, the condition that $L$ be time-like and $P$ space-like together with the reality conditions implies $W \bar{W} + Z \bar{Z}< 1$. This gives a complete description of the two metrics in terms of twistor data. We can now use this information to deduce a complex structure and Przanowski's function in both cases.\\
We set $l^0 = 0$ to select a holomorphic surface in $T$, from \eqref{CP2TL} we see that this amounts to choosing the complex structure induced from $S_1 = \{ (l,p) \in T \; | \; ( \xi^{0'}, \xi^{1'}) = (1,0) \}$. The twistor lines restricted to $S_1$ are

\begin{align*}
 l^j = \left( 0,1,-Z \right), \quad \quad \quad
 p_j = \left(1+Z\tilde{Z}, -\tilde{W}Z, -\tilde{W} \right), 
\end{align*}
so we can choose holomorphic coordinates $z = Z$ and $w = \frac{(1+Z\tilde{Z})}{W\tilde{W}}W$. Note that the contact form restricted to $S_1$ is indeed proportional to $dz$:

\begin{align*}
 \tau\Bigg|_{S_1} = \frac{\tilde{W}}{\left( 1+W\tilde{W}+Z\tilde{Z} \right) \left( 1+Z \tilde{Z} \right)^2} dZ.
\end{align*}
The parametrisation \eqref{CP2TL} is chosen to ensure that $\iota(S_1) = S_2$ where $S_2 = \{ (l,p) \in T \; | \; ( \xi^{0'}, \xi^{1'}) = (0,1) \}$ for both reality conditions. On $S_2$ we have

\begin{align*}
 l^j = \left( -(1+Z\tilde{Z}), W\tilde{Z},W \right), \quad \quad \quad
 p_j = \left( 0, 1, -\tilde{Z} \right),
\end{align*}
so we can choose anti-holomorphic coordinates $\zbar = - \epsilon \tilde{Z}$ and $\wbar = - \epsilon \frac{(1+Z\tilde{Z})}{W\tilde{W}}\tilde{W}$. Again

\begin{align*}
 \tau\Bigg|_{S_2} = W d\tilde{Z}
\end{align*}
as required. To retrieve the Przanowski function we need only use \eqref{K},

\begin{align*}
 K = \frac{1}{\Lambda} \ln \left[ \frac{W \tilde{W}}{ \left( 1+W\tilde{W}+Z\tilde{Z} \right) \left( 1+Z\tilde{Z} \right)^2}\right],
\end{align*}
which is valid for arbitrary cosmological constant. In terms of the coordinates $(w,z, \wbar, \zbar)$ and taking account of reality conditions we have

\begin{align*}
 K = -\frac{1}{\Lambda} \ln \left[ \left( 1 - \epsilon w \wbar - \epsilon z \zbar \right) \left( z\zbar - \epsilon \right) \right].
\end{align*}

\section{Conclusion}
We considered Quaternion-K\"ahler four-manifolds, which by definition are anti-self-dual Einstein. We introduced their local description by Przanowski's function $K$ and showed that metrics of this form are anti-self-dual Einstein provided $K$ satisfies Przanowski's equation \eqref{Prz}.\\
We continued with twistorial techniques to construct a Lax Pair, i.e. two vector fields $l_A$ that commute if and only if Przanowski's equation is satisfied. The existence of this Lax Pair confirms that Przanowski's equation is integrable, as one would expect from an equation coming from self-duality equations in four dimensions.\\
Furthermore, we encountered a conformally invariant differential operator acting on the line bundle $L^{l,m}$ as well as recursion relations relating solutions of the associated Laplace equation to cohomology classes on twistor space. Special cases are the conformal Laplacian and the linearised Przanowski operator. The latter annihilates perturbations $\delta K$ of Przanowski's function and thus describes deformations of the underlying manifold. We explained how the corresponding deformation of the twistor data is determined by the associated cohomology class $H^1 ( T, \mathcal{O}(2) )$. We also constructed an contour integral formula for $\delta K$ in terms of this cohomology class. If desired, it would be straight forward to write down a contour integral for all other values of $(n,k)$.\\
The next section was dedicated to the procedure of recovering a complex structure and Przanowski's function $K$ with the associated choice of holomorphic coordinates from twistor data. We illustrated the necessary steps explicitly using a number of examples including the non-trivial cases of $\C P^2$ with the Fubini-Study and $\widetilde{\C P^2}$ with the Bergmann metric. The latter is an interesting starting point for deformations, as $\C P^2$ with the Fubini-Study metric is rigid.\\
Looking beyond the four-dimensional case, it would be interesting to see how much of the local description of a Quaternion-K\"ahler metric by a scalar function with one associated second-order partial differential equation remains valid in higher dimensions. Some comments in this direction have been made in \cite{Alexandrov:2009vj} and some rigorous claims appear in \cite{Ogievetsky}, however in a much more physical setup.\\

\section*{Acknowledgements}
I am grateful to my supervisor Maciej Dunajski for invaluable input throughout the progress of this project. Also, I would like to thank Stefan Vandoren and Martin Wolf for very helpful discussions.

\section*{Appendix A}
We show that Przanowski's equation \eqref{Prz} is both sufficient and necessary for $[l_0, l_1]=0$, where $l_A$ are given by \eqref{LaxPair}. Writing the various components of the commutator as

\begin{align*}
 [l_0,l_1] = A \nabla_{01'} + B \nabla_{11'} + \left( C_1 \xi + C_2 \xi^2 + C_3 \xi^3 \right) \p_{\xi},
\end{align*}
we find

\begin{align*}
 A &= \frac{\xi \Kw \Kzw - \xi^2 \Kwbar}{\tilde{K} \Kw \Kwbar} \left( \tilde{K} + \Kw \Kwbar e^{\Lambda K} \right), \quad \quad \quad \quad \quad B = -  \frac{\xi \Kww}{\tilde{K} \Kw} \left( \tilde{K} + \Kw \Kwbar e^{\Lambda K} \right),\\
 C_1 &= - \frac{1}{\Kwbar} \left(\Kww \p_z - \Kzw \p_w \right) \left( \frac{e^{\Lambda K} \Kw \Kwbar}{\tilde{K}} \right), \quad \quad \quad \quad \; \;
 C_3 = - \frac{1}{\Kw} \nabla_{01'} \left( \frac{e^{\Lambda K} \Kw \Kwbar}{\tilde{K}} \right), \\
 C_2 &= \left( \nabla_{00'} \nabla_{11'} - \nabla_{10'} \nabla_{01'} + \frac{e^{\Lambda K} \Kwbar}{\tilde{K}} \p_w + \frac{e^{\Lambda K} \Kw}{\tilde{K}} \p_{\wbar}  + \frac{2 e^{\Lambda K} \Kww}{\tilde{K}} \right) \left( 1- \frac{\tilde{K}}{e^{\Lambda K} \Kw \Kwbar} \right).
\end{align*}
First consider the coefficient $A$, it vanishes only if the Przanowski equation \eqref{Prz} holds. Conversely, if \eqref{Prz} is satisfied all coefficients vanish.

\section*{Appendix B}
Following \cite{Mason, Dunajski2009} we introduce spinor formalism in four dimensions. Under the group isomorphism $SO(4,\C) \cong SL(2,\C) \times SL(2,\C)$ the tangent bundle $T \M$ of a four-dimensional complexified Riemannian manifold $(\M,g)$ with holomorphic metric $g$ can be regarded as a tensor product $T \M = \spS \otimes \spS'$ of two rank 2 spin bundles $\spS$ and $\spS'$. We can choose a null tetrad $e^{AA'}$ of $T^*\M$, in which

\begin{align}
\label{notnulltetrad}
 g = \varepsilon_{AB} \varepsilon_{A'B'} e^{AA'} \; e^{BB'} =  2 \left( e^{00'} e^{11'} - e^{01'} e^{10'} \right).
\end{align}
Primed and unprimed indices will always run from 0 to 1. Equation \eqref{notnulltetrad} amounts to choosing a basis $(o^A,\rho^A)$ of $\spS$ and a basis $(o^{A'}, \rho^{A'})$ of $\spS'$ over every point of $\M$ and setting

\begin{align*}
 e^{00'} = o_A o_{A'} e^{AA'}, \quad \quad  e^{01'} = o_A \rho_{A'} e^{AA'}, \quad \quad  e^{10'} = \rho_A o_{A'} e^{AA'}, \quad \quad  e^{11'} = \rho_A \rho_{A'} e^{AA'}.
\end{align*}
The metric thus induces symplectic structures $\epsilon_{AB}$ on $\spS$ and $\epsilon_{A'B'}$ on $\spS'$, which in the basis $(o,\rho)$ and $(o', \rho')$ are simply given by the Levi-Civita symbols $\varepsilon_{AB}$, $\varepsilon_{A'B'}$.\\
On the Lie algebra level, we have an induced isomorphism of $so(4,\C) \cong sl(2,\C) \oplus sl(2,\C)$ which leads to a splitting of the Levi-Civita connection $\Gamma$. Taking Cartan's first structural equation\footnote{We are suppressing the one-form index.}

\begin{align*}
 d  e^{AA'}  = {\Gamma^{AA'}}_{CC'} \wedge e^{CC'}
\end{align*}
as the definition of the connection coefficients and writing

\begin{align*}
 \Gamma_{AA'CC'} = \epsilon_{AC} \Gamma_{A'C'} + \epsilon_{A'C'} \Gamma_{AC},
\end{align*}
we find that this splitting is realised by the decomposition of $\Gamma$ into a symmetric unprimed connection $\Gamma_{AB}$ on $\spS$ and a symmetric primed connection $\Gamma_{A'B'}$ on $\spS'$. Similarly, the curvature of $\M$ splits up into the primed curvature ${R^{A'}}_{B'}$ of $\spS'$ and the unprimed curvature ${R^A}_B$ of $\spS$, where

\begin{align*}
 {R^A}_B = d {\Gamma^A}_B + {\Gamma^A}_C \wedge {\Gamma^C}_B, \quad \quad \quad {R^{A'}}_{B'} = d {\Gamma^{A'}}_{B'} + {\Gamma^{A'}}_{C'} \wedge {\Gamma^{C'}}_{B'}.
\end{align*}
Lastly, we can define a basis of the self-dual two-forms $\Sigma^{A'B'}$ as well as the anti-self-dual two-forms $\Sigma^{AB}$ by

\begin{align*}
  \Sigma^{A'B'} = \frac{1}{2} \epsilon_{AB} e^{AA'} \wedge e^{BB'}, \quad \quad \quad
  \Sigma^{AB} = \frac{1}{2} \epsilon_{A'B'} e^{AA'} \wedge e^{BB'}.
\end{align*}
These satisfy identities

\begin{align}
\label{SigmaID}
 \Sigma^{A'B'} \wedge \Sigma^{C'D'} = \frac{1}{4} \epsilon_{AB} \epsilon_{CD} e^{AA'} \wedge e^{BB'} \wedge e^{CC'} \wedge e^{DD'},
\end{align}
and similarly for $\Sigma^{AB}$, while all other wedge products vanish. Using these two-forms to decompose the primed and unprimed curvature spinors, we find

\begin{align*}
 {R^{A'}}_{B'} &= \frac{1}{12} R {\Sigma^{A'}}_{B'} + {W^{A'}}_{B'C'D'} \Sigma^{C'D'} + {\varPhi^{A'}}_{B'CD} \Sigma^{CD},\\
 {R^A}_B &= \frac{1}{12} R {\Sigma^A}_B + {W^A}_{BCD} \Sigma^{CD} + {\varPhi^A}_{BC'D'} \Sigma^{C'D'}.
\end{align*}
Here ${W^A}_{BCD}$ and ${W^{A'}}_{B'C'D'}$ are the anti-self-dual and self-dual Weyl spinor, ${\varPhi^A}_{BC'D'}$ is the trace-free Ricci spinor and $R = 12 \Lambda$ is the scalar curvature.

\vfill

\pagebreak

\end{document}